\documentclass[twocolumn,showpacs,preprintnumbers,showkeys,superscriptaddress,prc]{revtex4}
\usepackage{CJK}
\usepackage{mathrsfs}
\usepackage{amssymb}
\usepackage{amsmath}
\usepackage{graphicx}
\usepackage{dcolumn}
\usepackage{bm}
\usepackage{graphicx}
\usepackage{float}
\usepackage[normalem]{ulem}
\usepackage{color}
\usepackage{multirow}

\newcolumntype{d}[1]{Dc{.}{.}{#1}}
\begin{document}
\begin{CJK*}{UTF8}{} 

\title{random-interaction study on linear systematics of $I^{\pi}=11/2^-$ electromagnetic moments in Cd isotope chain}

\author{Z. Z. Qin ({\CJKfamily{gbsn}覃珍珍})}
\affiliation{School of Science, Southwest University of Science and Technology, Mianyang 621010, China}
\author{Y. Lei ({\CJKfamily{gbsn}雷杨})\footnote{corresponding author: leiyang19850228@gmail.com}}
\affiliation{Key Laboratory of Neutron Physics, Institute of Nuclear Physics and Chemistry, China Academy of Engineering Physics, Mianyang 621900, China}
\date{\today}

\begin{abstract}
In the random-interaction ensembles, electromagnetic moments of Cd $I^{\pi}=11/2^-$ isomers predominately present linear systematics as changing the neutron number, which has been reported in realistic nuclear system. Quadrupole-like and $\delta$-like $pn$ interaction are responsible for such linear systematics of quadrupole and magnetic moments, respectively.
\end{abstract}
\maketitle
\end{CJK*}

\section{interaction}
The low-lying spectra and magnetic properties of doubly even nuclei are highly regulated with simple patterns. For examples, they always have $I^{\pi}=0^+$ ground states, and $I^{\pi}=2^+$ second excited states with few exceptions; their quadrupole moments of $2^+_1$ and $2^+_2$ states generally present a strong correlation with similar magnitudes and different signs across the whole nuclide chart \cite{allmond}. On the other hand, these two regularities robustly exist in an ensemble of nuclear models with random numbers as two-body interaction matrix elements \cite{johnson-prl,q1q2-lei}. These two and other robust regularities in the random-interaction ensemble demonstrates how {\it simple} regularity emerges out of {\it complex} nuclear system, even with interactions mostly deviating from reality \cite{rand-rev-1,rand-rev-2,rand-rev-3,rand-rev-4,rand-book}.

Recently, it was reported that the $I^{\pi}=11/2^-$ electromagnetic moments of neutron rich Cd isotopes are also simply regulated with an obvious linear systematics as changing neutron number \cite{cd11}. Several theoretical investigations tried to understand this linear systematics based on the BCS \cite{cd11-bcs}, density functional theory \cite{cd11-zhao}, and schematic Shell Model \cite{cd11-lei}. However, it's still the challenge to explain how such {\it simplicity} survives out of the {\it complex} nuclear structure \cite{cd11-wood}. We believe that the random-interaction ensemble may provide some clue. Furthermore, most previous random-interaction works focused on the robust properties of a single doubly even nucleus. It's novel to apply random interaction to a nuclear-systematics study on an odd-mass isotope chain. Therefore, our work aims to probe and understand the robustness of such linear systematics in the random-interaction ensemble.

\section{calculation framework}\label{cal}
Our random-interaction calculations covers $^{112\sim 130}$Cd, whose single-particle orbits are limited to $\pi 0g{9/2}$, $\nu 2s_{1/2}$, $\nu 1d{3/2}$ and $\nu 0h_{11/2}$ within $Z=40\sim 50$ and $N=64\sim 82$ shell closures. No further truncation is introduced. Single-neutron energies are set to be degenerated, almost as reality. The two-body interaction is randomized within the two-body random ensemble (TBRE) \cite{tbre-1,tbre-2,tbre-3}. In other words, any two-body interaction element denoted by $V_{j_1j_2j_3j_4}^J$ follows the Gaussian distribution with $(\mu=0,~\sigma^2=1+\delta_{j_1j_2,j_3j_4})$, where $j_1$, $j_2$, $j_3$ and $j_4$ represent the four single-particle orbits, and the superscript $J$ labels the rank.

We generate 3 000 000 sets of two-body interaction elements, and input them into the shell-model code \cite{sm-code}. With each set of two-body interaction elements, we first calculate corresponding low-lying spectra of even-mass Cd. If the calculation produces $I^{\pi}=0^+$ ground states for all the even-mass Cd isotopes, we further calculate the $I^{\pi}=11/2^-$ electromagnetic moments of odd-mass Cd with the same set of elements. To simplify our description, these quadrupole and magnetic moments are denoted by $Q$ and $\mu$, respectively. For $Q$ calculations, effective charges are set as $e_{\pi}=1.5$e and $e_{\nu}=0.5$e; For $\mu$ calculations, single-particle $g$ factors are set as $g_{\pi s}=5.586\times0.7\mu_N$, $g_{\pi l}=1\mu_N$, $g_{\nu s}=-3.826\times0.7\mu_N$ and $g_{\nu l}=0$, where the spin $g$ factors are conventionally quenched by 0.7.

To quantitatively describe the $Q$ and $\mu$ systematics, we introduce the Pearson correlation coefficient (denoted by $\rho$) \cite{pcc} as a measure of linear correlation between electromagnetic moment and neutron number. This coefficient has a value between $\pm 1$, where 1, 0 and -1 correspond to totally positive linear correlation, no linear correlation, and totally negative linear correlation, respectively. For instance, according to Table I of Ref. \cite{cd11}, the experimental $Q$ values present $\rho=0.997$ linear systematics, and $\mu$ values present $\rho=0.862$ smaller than $Q$. Thus, the $Q$ linearity is more evident than $\mu$ one as observed.

\section{interaction property with $I^{\pi}=0^+$ ground states}\label{gs}
\begin{figure}
\includegraphics[angle=0,width=0.45\textwidth]{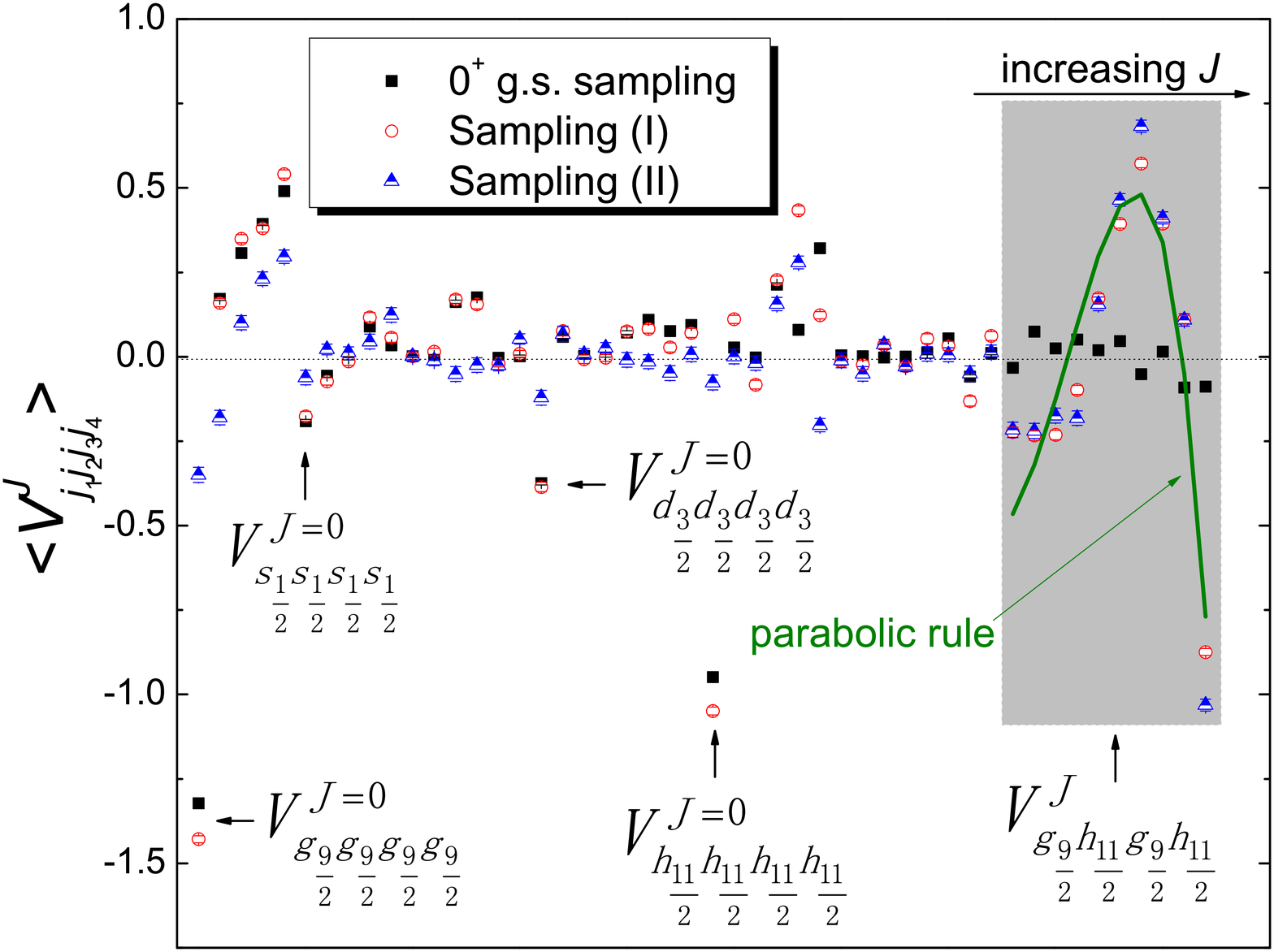}
\caption{(Color online) $\langle V^J_{j_1j_2j_3j_4}\rangle$ values after the $0^+$-g.s. sampling, sampling (I) and sampling (II) in an ascending order of $j_1j_2j_3j_4J$ indexes. Sampling (I) corresponds to the sampling with $\rho>0.9$ $Q$ systematics and $0^+$ ground states for all the even-mass Cd isotopes. Sampling (II) corresponds to the sampling with $\rho>0.9$ $Q$ systematics regardless of even-mass Cd ground states. The grey areas highlight the proton-neutron interaction between $\pi 0g_{9/2}$ and $\nu 0h_{11/2}$, where the evolution with the parabolic rule \cite{paar-pr} is schematically illustrated with a solid curve. The dash line illustrates the ensemble mean of the TBRE. The error bar is determined by statistic error.}\label{q11-int}
\end{figure}

Our $Q$ and $\mu$ calculations are based on the interactions, which can provide $I^{\pi}=0^+$ ground states (denoted by $0^+$ g.s.) for all the even-mass Cd isotopes. In other words, we perform a sampling in the TBRE before the $Q$ and $\mu$ calculations. Only $\sim$1\% interactions can survive such sampling, although $0^+$-g.s. predominance is still preserved for each single even-mass Cd isotope in the TBRE \cite{johnson-prl}. We present $V^J_{j_1j_2j_3j_4}$ average values (denoted by $\langle V^J_{j_1j_2j_3j_4}\rangle$) after this $0^+$-g.s. sampling as the background of our analysis in Fig. \ref{q11-int}. 


\begin{figure}
\includegraphics[angle=0,width=0.45\textwidth]{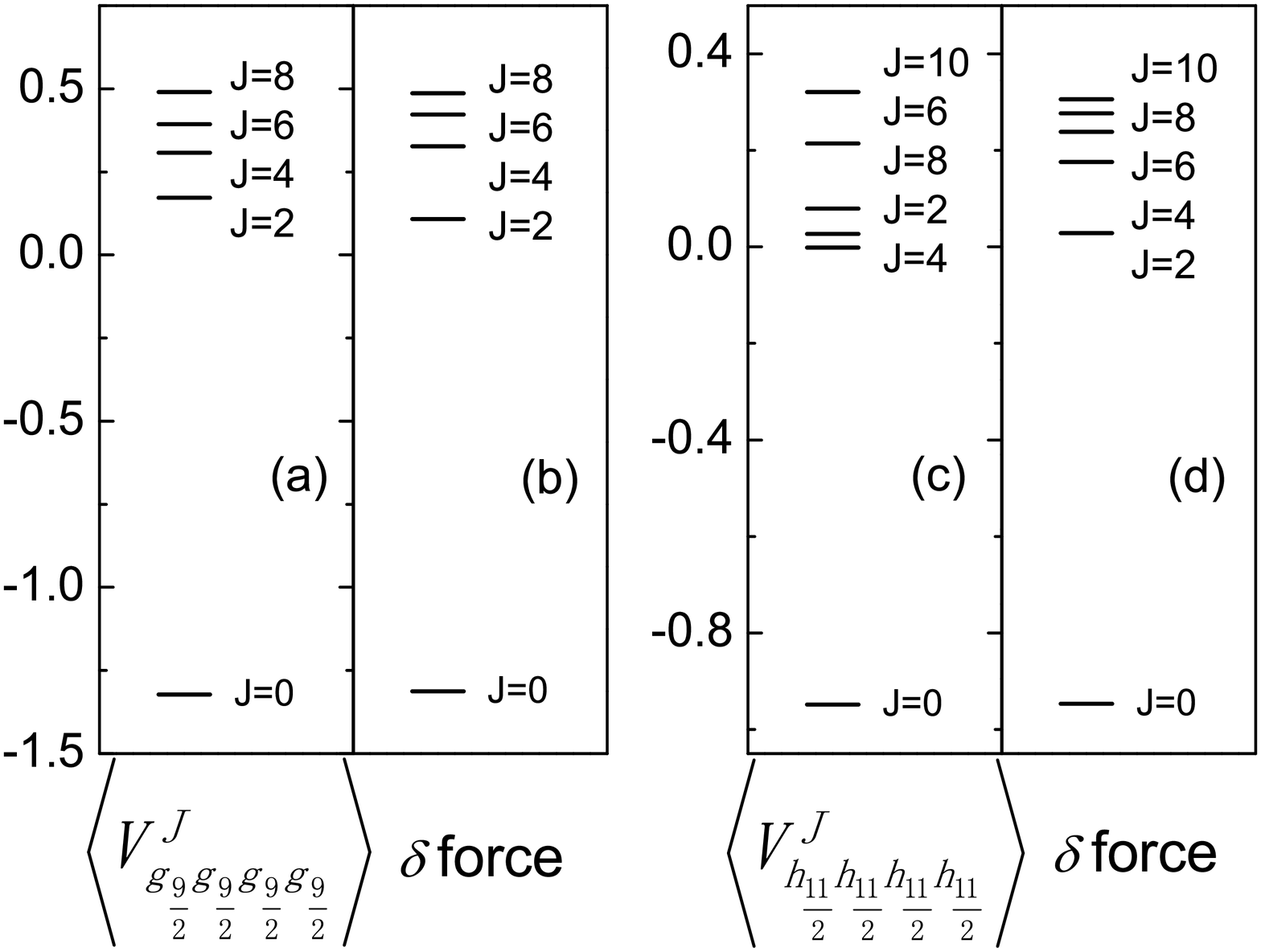}
\caption{Comparison between $\langle V^J_{j_1j_2j_3j_4}\rangle$ values after the $0^+$-g.s. sampling and the $\delta$-force elements.}\label{delta}
\end{figure}

According to Fig. \ref{q11-int}, all the $\langle V^J_{j_1j_2j_3j_4}\rangle$ values with $J=0$ after the $0^+$-g.s. sampling are relatively more attractive than others, corresponding to the short-range property of nuclear force in realistic nuclear system. To visualize this more straightforwardly, we compare the $\langle V^J_{g_{\frac{9}{2}}g_{\frac{9}{2}}g_{\frac{9}{2}}g_{\frac{9}{2}}}\rangle$ and $\langle V^J_{h_{\frac{11}{2}}h_{\frac{11}{2}}h_{\frac{11}{2}}h_{\frac{11}{2}}}\rangle$ values with two-body interactions elements of a typical short-range interaction, {\it i.e.} the $\delta$ force, in Fig. \ref{delta}. The similarity between them is obvious. Therefore, we conclude that even through the interaction origin of the $0^+$-g.s. predominance for a single nucleus in the TBRE is unclear, the short-range property of nuclear force is still the key to keep all the doubly even nuclei have $0^+$ g.s. in both TBRE and realistic nuclear system.

\section{$Q$ linearity}\label{sec-q}

\begin{figure}
\includegraphics[angle=0,width=0.45\textwidth]{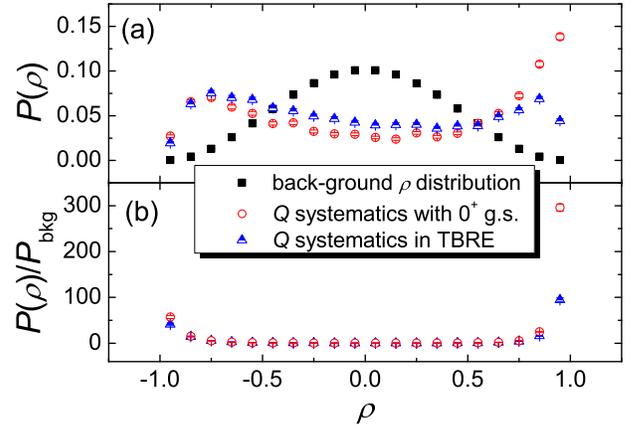}
\caption{(Color online) $\rho$ distributions of the $Q$ systematics after the $0^+$-g.s. sampling and in the whole TBRE compared with the back-ground distribution (denoted by $P_{\rm bkg}$, see text for definition). The error bar is determined by statistic error.}\label{q11-dis}
\end{figure}

After the $0^+$-g.s. sampling, we calculate the $\rho$ distribution of the $Q$ systematics as presented in Fig. \ref{q11-dis}(a), where the predominance with $P(\rho>0.9)\simeq 15$\% emerges, corresponding to the almost linear $Q$ evolution in the realistic nuclear system. However, there is still considerable distribution of $P(\rho\sim 0)\simeq 5$\%, corresponding to totally chaotic $Q$ evolutions. These chaotic evolutions should be taken as the back ground of our analysis, because the TBRE intrinsically and trivially maintains chaos. To illustrate the back ground, we generate 25 000 000 sets of independently randomized $Q$ values from the normal distribution instead of from the TBRE, and present their $\rho$ distribution as our back-ground distribution (denoted by $P_{\rm bkg}$) in Fig. \ref{q11-dis}(a). To eliminate this interference from the back-ground distribution, we present the ratio of $P(\rho)$ for the $Q$ systematics over $P_{\rm bkg}$ in Fig. \ref{q11-dis}(b), where the predominance of the $Q$ linearity is more obvious with $P(\rho>0.9)/P_{\rm bkg}\simeq 300$.

One may argue that, the linear $Q$ systematics out of the $0^+$-g.s. sampling is trivial, because the $0^+$-g.s. sampling favors the $\delta$-like interaction, and thus should enhance the seniority scheme, which has been proposed to be the origin of the linearity of $Q$ systematics in Ref. \cite{cd11}. To examine this argument, we also calculate the $\rho$ distribution of the $Q$ systematics in the whole TBRE without the $0^+$-g.s. sampling as shown in Fig. \ref{q11-dis}. $P(\rho>0.9)$ with the $0^+$-g.s. sampling is 3 times of that without this sampling according to Fig. \ref{q11-dis}(a), which agrees with the claim in Ref. \cite{cd11}, that the seniority scheme indeed enhances the linearity of the $Q$ systematics. However, in Fig. \ref{q11-dis}(b), the predominance of $P(\rho>0.9)/P_{\rm bkg}$ is still obvious, even without the $0^+$-g.s. sampling. This means that the $Q$ linearity is robust in the whole TBRE, which can not be totally attributed to the seniority scheme here.

To search the origin of this $Q$ linearity in the TBRE, we perform two additional samplings: 
\begin{itemize}
\item[(I)]
the sampling with $\rho>0.9$ and $0^+$ ground states for all the even-mass Cd isotopes; 
\item[(II)]
the sampling with $\rho>0.9$ regardless of even-mass Cd ground states.
\end{itemize}
The $\langle V^J_{j_1j_2j_3j_4}\rangle$ values of these two samples are presented in Fig. \ref{q11-int} compared with those after the $0^+$-g.s. sampling. Sampling (I) and the $0^+$-g.s. sampling share the same short-range property of like-nucleons interaction, {\it i.e.}, relatively attractive interaction elements with rank $J=0$, which can be trivial, because sampling (I) is actually based on the $0^+$-g.s. sampling. Furthermore, after sampling (I), the proton-neutron ($pn$) interaction elements between $\pi 0g_{9/2}$ and $\nu 0h_{11/2}$ orbits obviously follow the parabolic rule \cite{paar-pr} as increasing rank $J$, corresponding to the quadrupole interaction \cite{casten-book}. Sampling (II) also favors a quadrupole-like $pn$ interaction, even though the rank-$J=0$ interaction elements after this sampling present not short-range property. Therefore, we conclude that the quadrupole $pn$ interaction is responsible to induce the $Q$ linearity, and the seniority scheme is a boost to this linearity in random-interaction ensemble.

\section{$\mu$ linearity}\label{sec-mu}

\begin{figure}
\includegraphics[angle=0,width=0.45\textwidth]{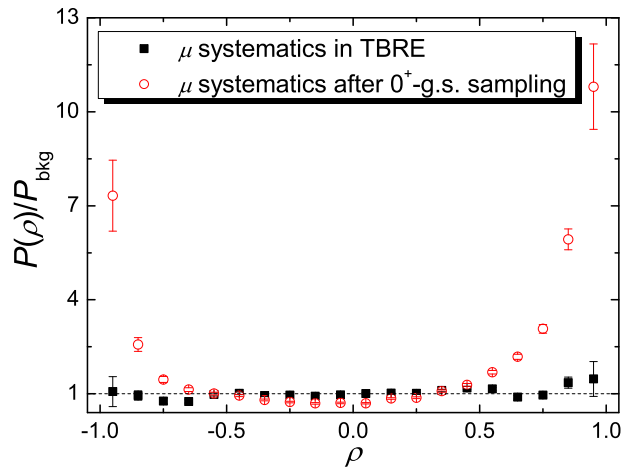}
\caption{(Color online) $\rho$ distributions of the $mu$ systematics in the whole TBRE and with the $0^+$-g.s. sampling, normalized with $P_{\rm bkg}$. The dash line with $P(\rho)/P_{\rm bkg}=1$ highlights a totally independent $\mu$ variation  from neutron number. The error bar is determined by statistic error.}\label{mu11-dis}
\end{figure}

In Fig. \ref{mu11-dis} we present $\rho$ distributions of the $\mu$ systematics with the $0^+$-g.s. sampling and in the whole TBRE, normalized with $P_{\rm bkg}$. Here $P_{\rm bkg}$ for $\mu$ systematics should be the same as that for $Q$. In the TBRE, $P(\rho)/P_{\rm bkg}$ is always close to 1, which demonstrates that the TBRE does not characterize the $\mu$ systematics. However, after the $0^+$-g.s. sampling, relatively larger possibility for $|\rho|>0.9$ emerges ($\sim$1\% in the $0^+$-g.s. sample), corresponding to the the predominance of the $\mu$ linearity.

\begin{figure}
\includegraphics[angle=0,width=0.45\textwidth]{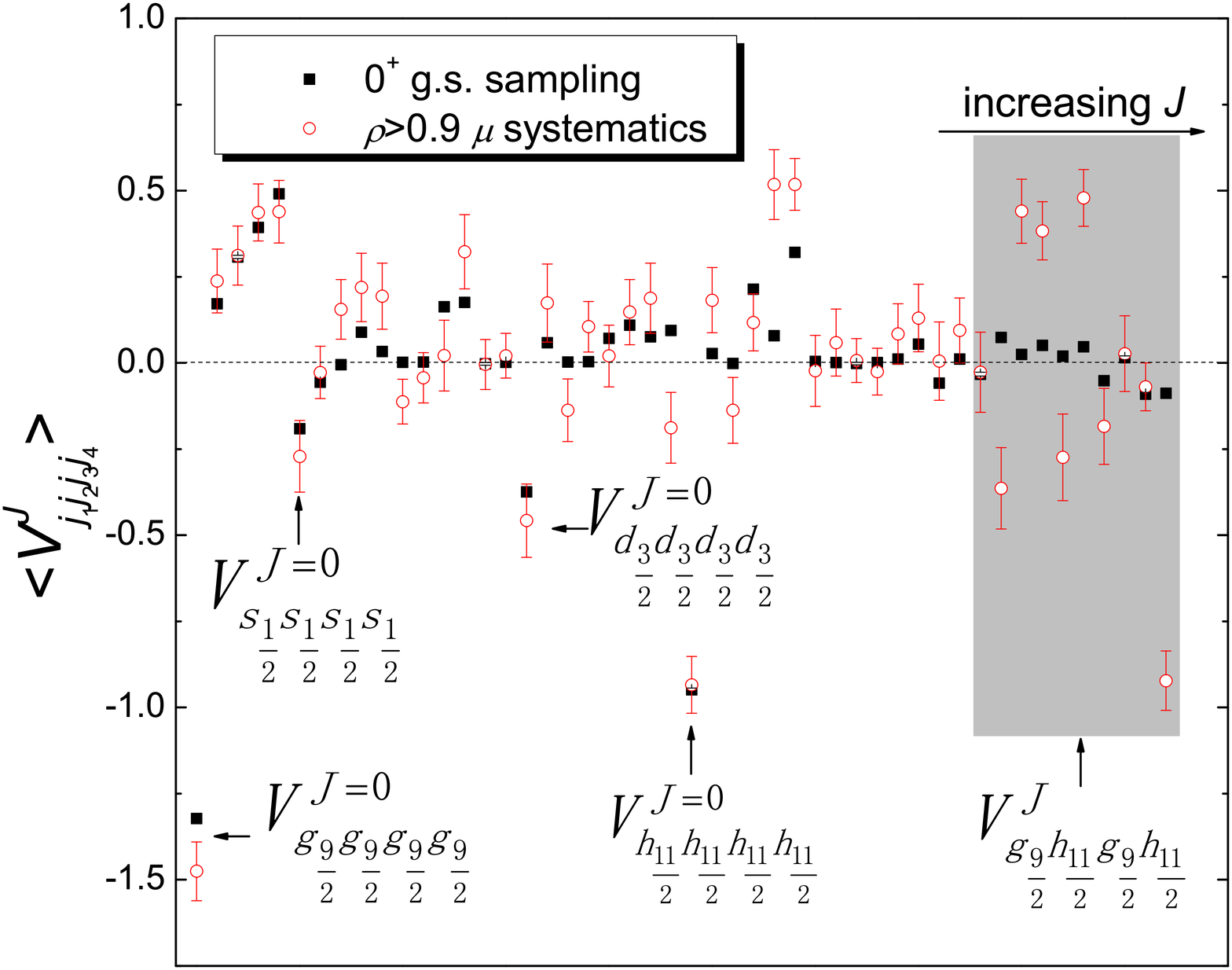}
\caption{(Color online) $\langle V^J_{j_1j_2j_3j_4}\rangle$ values after the $0^+$-g.s. sampling and the $\rho>0.9$ sampling of $\mu$ systematics beyond the $0^+$-g.s. sampling, in an ascending order of $j_1j_2j_3j_4J$ indexes. The grey areas highlight characteristic of the $pn$ interaction elements surviving the $\rho>0.9$ sampling. The dash line illustrates the ensemble mean of the TBRE. The error bar is determined by statistic error.}\label{mu11-int}
\end{figure}

It seems that the $\mu$ linearity requires even-mass Cd $0^+$ ground states, {\it i.e.} the seniority scheme as we have explained. However, pure seniority scheme can only provide a constant $\mu$ as agued in Ref. \cite{cd11}. Thus, we need to further probe other origin of the $\mu$ linearity beside the seniority scheme. We perform a sampling for $\rho>0.9$ $\mu$ systematics based on the $0^+$-g.s. sampling, and present $\langle V^{J}_{j_1j_2j_3j_4}\rangle$ values after such sampling in Fig. \ref{mu11-int}. $\rho<-0.9$ linearity is totally against experiments, and thus omitted here.

\begin{figure}
\includegraphics[angle=0,width=0.45\textwidth]{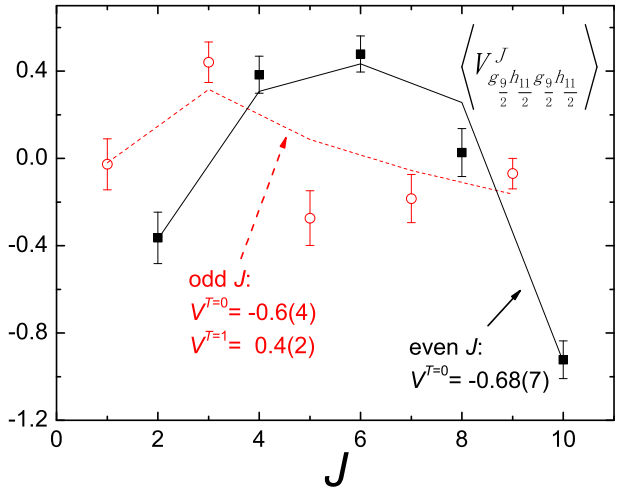}
\caption{(Color online) $\langle V^J_{g_{9/2}h_{11/2}g_{9/2}h_{11/2}}\rangle$ after the $\rho>0.9$ sampling of $\mu$ systematics with all the even Cd having $0^+$ ground states. The even-$J$ behavior of $\langle V^J_{g_{9/2}h_{11/2}g_{9/2}h_{11/2}}\rangle$ is different from odd-$J$ one, and thus they are presented separately. Straight lines illustrates the fitting to $pn$ interaction elements governed by the $\delta$ force. The error bar is determined by statistic error.}\label{mu11-fit}
\end{figure}

In Fig. \ref{mu11-int}, the $\rho>0.9$ sampling and the $0^+$-g.s. sampling share the short-range property of the like-nucleon interaction, which means that the seniority scheme is still important for the $\mu$ linearity. Furthermore, the $\rho>0.9$ sampling presents additional structure of $\langle V^J_{g_{\frac{9}{2}}h_{\frac{11}{2}}g_{\frac{9}{2}}h_{\frac{11}{2}}}\rangle$. We replot the detail of $\langle V^J_{g_{\frac{9}{2}}h_{\frac{11}{2}}g_{\frac{9}{2}}h_{\frac{11}{2}}}\rangle$ of the $\rho>0.9$ sample in Fig. \ref{mu11-fit}. The even-$J$ behavior of $\langle V^J_{g_{\frac{9}{2}}h_{\frac{11}{2}}g_{\frac{9}{2}}h_{\frac{11}{2}}}\rangle$ is different from odd-$J$ one: the even-$J$ $\langle V^J_{g_{\frac{9}{2}}h_{\frac{11}{2}}g_{\frac{9}{2}}h_{\frac{11}{2}}}\rangle$ values present an obvious parabolic evolution, while those with odd $J$ seem less regulated. This odd-even difference also characterizes the $pn$ interaction governed by the $\delta$ force \cite{casten-book}. More specifically, the evolution of even-$J$ interaction between $\pi 0g_{9/2}$ and $\nu 0h_{11/2}$ orbit are only attributed to the $T=0$ $\delta$ force as
\begin{equation}\label{eq-even}
\begin{aligned}
V^{J={\rm even}}_{g_{\frac{9}{2}}h_{\frac{11}{2}}g_{\frac{9}{2}}h_{\frac{11}{2}}}&\propto
\begin{pmatrix}
9/2&11/2&J\\
1/2&-1/2&0
\end{pmatrix}
^2\\
&\times V^{T=0}\left\{1+\frac{121}{J(J+1)}\right\},
\end{aligned}
\end{equation} 
while the odd-$J$ $\delta$ interaction elements have both $T=0$ and $T=1$ contributions as 
\begin{equation}\label{eq-odd}
\begin{aligned}
V^{J={\rm odd}}_{g_{\frac{9}{2}}h_{\frac{11}{2}}g_{\frac{9}{2}}h_{\frac{11}{2}}}&\propto
\begin{pmatrix}
9/2&11/2&J\\
1/2&-1/2&0
\end{pmatrix}
^2\\
&\times \left\{ V^{T=1}+V^{T=0}\frac{1}{J(J+1)}\right\}.
\end{aligned}
\end{equation}

With $V^{T=0}$ and $V^{T=1}$ as fitting variables, we fit even-$J$ and odd-$J$ $\langle V^J_{g_{\frac{9}{2}}h_{\frac{11}{2}}g_{\frac{9}{2}}h_{\frac{11}{2}}}\rangle$ values to Eqs. (\ref{eq-even}) and (\ref{eq-odd}), respectively. For even $J$, the best-fit $V^{T=0}$=-0.68(7); for odd $J$, the best-fit $V^{T=0}$=-0.6(4) and $V^{T=1}$=0.4(2). The fitting error for odd $J$ is larger than that for even $J$. This is because the odd-$J$ fitting involves one more fitting variable. We also note that the best-fit $V^{T=0}$ values of even $J$ and odd $J$ agree with each other, which implies that $\langle V^J_{g_{\frac{9}{2}}h_{\frac{11}{2}}g_{\frac{9}{2}}h_{\frac{11}{2}}}\rangle$ for $\rho>0.9$ $\mu$ systematics are describable with a unified $\delta$ $pn$ interaction. According to the best-fit $V^{T=0}$ and $V^{T=1}$ values, to induce a linear $\mu$ evolution, the $T=0$ $pn$ interaction shall be attractive, while the $T=1$ one is repulsive. This has also been long noted in the realistic nuclear system, which may explain the linearity of experimental $\mu$ data.

\section{summary}
To summarize, the random-interaction ensemble predominantly reproduces the linear $Q$ and $\mu$ systematics in the Cd isotopes chain. The $pn$ interaction is the key to linearize the the $Q$ and $\mu$ systematics, although the seniority scheme is a significant boost. For the $Q$ linearity, the $pn$ interaction presents quadrupole-like feature. For the $\mu$ linearity, the $\delta$-like $pn$ interaction is required with repulsive $T=1$ and attractive $T=0$ components.

Our work also emphasizes that the short-range interaction between like nucleons is responsible to reproduce the $I^{\pi}=0^+$ ground states for all the even-mass nuclei in both TBRE and realistic nuclear system, which may provides a new viewpoint to understand the predominance of $I=0$ ground states in the TBRE.

\acknowledgements
This work was supported by the National Natural Science Foundation of China under Grant Nos. 11647059, 11305151, the Research Fund for the Doctoral Program of the Southwest University of Science and Technology under Grant No. 14zx7102, and the Graduate Education Reform Project of the Southwest University of Science and Technology under Grant No. 17sxb119.


\begin{thebibliography}{}
\bibitem{allmond} J. M. Allmond, Phys. Rev. C {\bf 88}, 041307 (2013).

\bibitem{johnson-prl} C. W. Johnson, G. F. Bertsch, and D. J. Dean, Phys. Rev. Lett. {\bf 80}, 2749 (1998).

\bibitem{q1q2-lei} Y. Lei, Phys. Rev. C {\bf 93}, 024319 (2016).

\bibitem{rand-rev-1} V. K. B. Kota, Phys. Rep. {\bf 347}, 223 (2001).

\bibitem{rand-rev-2} V. Zelevinsky and A. Volya, Phys. Rep. {\bf 391}, 311 (2004).

\bibitem{rand-rev-3} Y. M. Zhao, A. Arima, and N. Yoshinaga, Phys. Rep. {\bf 400}, 1 (2004). 

\bibitem{rand-rev-4} H. Weidenm\"{u}eller and G. E. Mitchell, Rev. Mod. Phys. {\bf 81}, 539 (2009).

\bibitem{rand-book} V. K. B. Kota, {\it Embedded Random Matrix Ensembles in Quantum Physics} (Springer, Heidelberg, 2014).

\bibitem{cd11} D. T. Yordanov, {\it et al.}, Phys. Rev. Lett. {\bf 110}, 192501 (2013).

\bibitem{cd11-bcs} N. B. de Takacsy, Phys. Rev. C {\bf 89}, 034301 (2014).

\bibitem{cd11-zhao} P. W. Zhao, S. Q. Zhang, and J. Meng, Phys. Rev. C {\bf 89}, 011301 (2014).

\bibitem{cd11-lei} Y. Lei, H. Jiang, and S. Pittel, Phys. Rec. C {\bf 92}, 024321 (2015).

\bibitem{cd11-wood} J. Wood, Physics 6, 52 (2013).

\bibitem{tbre-1} J. B. French and S. S. M. Wong, Phys. Lett. B {\bf 33}, 449 (1970).

\bibitem{tbre-2} O. Bohigas and J. Flores, Phys. Lett. B {\bf 34}, 261 (1971).

\bibitem{tbre-3} S. S. M. Wong and J. B. French, Nucl. Phys. A {\bf 198}, 188 (1972). 

\bibitem{sm-code} E. Caurier and F. Nowacki, Acta Phys. Pol. {\bf 30}, 705 (1999).

\bibitem{pcc} K. Pearson, Proc. R. Soc. London {\bf 58}, 240 (1895).

\bibitem{paar-pr} V. Paar, Nucl. Phys. A {\bf 331}, 16 (1979).

\bibitem{casten-book} R. F. Casten, in {\it Nuclear Structure From A Simple Perspective}, edited by P. E. Hodgson (Oxford University Press, 1990).

\end{thebibliography}
\end{document}